# The *Zhamanshin* Impact Event: Potential Implications for Environmental Responses and Biological Linkages on Earth and Beyond

James B. Garvin[1], Connor J. Anderson[2,3], Katherine A. Melocik[2,3], Devin R. McClain[2,3], Scott S. Sinno[2,3], Myoung-Jong Noh[4], and Compton J. Tucker[3,5]

Corresponding author: james.b.garvin@nasa.gov
[1] Sciences and Exploration Directorate, NASA Goddard Space Flight Center, 8800 Greenbelt Road, Greenbelt, MD 20771, USA
[2] Science Systems and Applications, Inc., 10210 Greenbelt Rd, Lanham, MD 20706, USA
[3] Earth Science Division, NASA Goddard Space Flight Center, 8800 Greenbelt Road, Greenbelt, MD 20771, USA
[4] Byrd Polar Research Center, The Ohio State University, 275A Scatt Hall, 1090 Carmack Rd, Columbus, OH 43210, USA
[5] University of Maryland Baltimore County, 1000 Hilltop Circle, Baltimore, MD 21250, USA

**Abstract**

At least one large-body (diameter > 1.1 km) hypervelocity cratering event occurred during ~ 0.8-0.90 Ma (*Zhamanshin,* Kazakhstan) in the Middle Pleistocene Transition period. Analysis designed to reduce uncertainty in the dimensions of the *Zhamanshin* structure employing high resolution topography demonstrated that it likely generated a ~ 26.5 km diameter multi-ring crater. This is at least two times larger than the current best estimates. Using a range of accepted impactor sizes, velocities, compositions, and angles of impact, such impacts typically yield kinetic energies of impact over 240,000 Megatons (TNT). Explosive energetic events of this magnitude (e.g., *Yellowstone Caldera*) at other times (K-Pg) have created global environmental effects. The factor of two discrepancy in the dimensions of *Zhamanshin* increases the kinetic energy yield by factors of 7-10, with significantly larger environmental consequences. This justifies examination of rapid climate transitions linked to biological consequences, including those related to environmental perturbations, at ~0.9 Ma.

## 1. INTRODUCTION

Exogenic events such as hypervelocity asteroid impacts on Earth have shaped the history of life (Gould, 2002; Grieve, 2017; Grieve and Shoemaker, 1994; Kring, 2003; Osinski et al., 2022; Pierazzo and Artemieva, 2012; Toon et al., 1997) with direct evidence from the *Chicxulub* impact at the K-Pg boundary (Morgan et al., 2022; Senel et al., 2023). Quantitative study of Earth's hypervelocity impacts is essential as they could have affected plant and animal populations amidst various environmental challenges during periods of known climate variations (e.g., Middle Pleistocene Transition: Clark et al., 2006). However, understanding the consequences of such events is challenging due to the significant uncertainty in the measurements of several key physical dimensions of terrestrial impact craters (Croft, 1985; Garvin et al., 1992; Turtle et al., 2005). These spatial dimensions, such as outer rim diameter (D), are related to a set of impactor parameters that are used to estimate the impact kinetic energy released during hypervelocity impact events (Collins et al., 2005; Davison and Collins, 2022; Osinski et al., 2022). Reducing the uncertainty of such physical dimensions of complex impact craters is imperative to better estimate the total kinetic energy released during terrestrial hypervelocity impact events and the environmental effects such large explosions could have on the global Earth ecosystem. Only ultra-recent and minimally eroded complex impact structures are well suited for this investigation given new recognition of erosional modification of apparent impact crater diameters under some circumstances (Ait Oufella et al., 2025).

Five complex impact cratering events are believed to have occurred within the last approximately one million years of Earth history, including: (1) *Iturralde*, Boliva; (2) *Pantasma*, Nicaragua; (3) *Zhamanshin*, Kazakhstan; (4) *Bosumtwi*, Ghana; *and* (5) *Muong Nong Tektite source crater*, Laos (Schmieder and Kring, 2020; Sieh et al., 2020). This study focuses on reassessing the geometric dimensions of the *Zhamanshin* complex impact structure, which is the largest and youngest of these with < 1 million years of arid-region erosion. Florenskii and Dabizha (1980) characterized this impact structure and its preserved morphology with subsequent studies focused on its state of geomorphic erosion (e.g., Burba, 1994). *Zhamanshin* is the youngest (< 0.90 Ma) and likely best-preserved complex impact structure known within the Earth record within the past one million years (Deino et al., 1990; Garvin et al., 2023; Garvin and Schnetzler, 1992). Garvin and Schnetzler (1992) discussed best available ~90 m horizontal scale digital topography (early 1990s) and concluded it was partially modified by erosion relative to the similarly sized impact features. The deposition of widespread loess atop the landscape further subdues its relief in comparison to other contemporaneous complex craters, such as *Bosumtwi* in Ghana (Wulf et al., 2019). Conjectures about why *Zhamanshin* displays little of its likely original topographic expression as a pristine multi-ringed complex impact structure have emphasized the unique physiography of the feature (Garvin et al., 1992; Garvin and Schnetzler, 1992; Indu et al., 2022), with the possibility of rapid modification-stage decay of initial cratering-related topography due to the combination of target effects and possible incursion of water from the nearby Aral Sea with associated fluvial runoff. Further investigation by Garvin and Schnetzler (1992) was limited to low vertical accuracy surface elevation data (~ 10-13 m) in this region of low-relief landforms (< 130 m). USSR geologists [Masaitis, Feld'man, Izokh *and others*], J. B. Garvin, and R. A. F. Grieve participated in a field expedition to *Zhamanshin* in Fall 1989 to study the surrounding landscape and to collect drill core samples for radiometric age determination (Deino et al., 1990). At that time, the international field team evaluated the possible outermost "tectonic rim" of the very recent impact structure, with tantalizing evidence of geomorphologic relief features suggestive of a true crater diameter greater than ~20 km (**Appendix 1** *Figures A1-A3*). However, the inadequacy of available digital topographic data with meters-scale vertical precision and limited gravimetry inhibited a fully quantitative analysis at that time.

Recent advancements in digital elevation model (DEM) data sets allow for an improved quantitative analysis of the *Zhamanshin* impact structure. DEMs with high spatial resolution and vertical accuracy can be produced at large geographic scales using stereo-photogrammetry (Howat et al., 2019; Porter et al., 2024). The use of a suite of new DEM data sets with high spatial and vertical resolution enables evaluation of critical dimensions at *Zhamanshin* given the regional expression of topography for the complex crater is subdued even after less than one million years of erosion. As such, we applied the Radial Profile Analysis System (Anderson et al., 2025) to five DEMs of varying spatial resolution and vertical precision to quantitatively identify possible topological "rings" (of detectable relief features) related to impact cratering mechanics for the *Zhamanshin* impact structure. The methodology used within the Radial Profile Analysis System was designed for terrestrial impact structures, and it was validated using five terrestrial impact craters of varying diameters and complexity with well-constrained dimensions in a recent study by Anderson *et al.* (2025) using the latest published dimensions (Osinski et al., 2022; Schmieder and Kring, 2020) (UWO Earth Impact Crater database accessible online at https://impact.uwo.ca/). Total impact kinetic energy and the associated environmental effects are predicted for the *Zhamanshin* impact event using the apparent outer rim dimensions computed via the Radial Profile Analysis System as inputs into the on-line Earth Impact Effects (EIE) assessment system (Collins et al., 2005).

## 2. METHODOLOGY

*2.1 Zhamanshin Impact Structure (Kazakhstan)*

The *Zhamanshin* impact structure was formed ~0.87 Ma in Kazakhstan roughly 200 km northwest of the Aral Sea (48° 21' 58" N; 60° 56' 9" E; **Figure 1**) (Florenskii and Dabizha, 1980). Recent research indicated a possible carbonaceous chondritic impactor (Magna et al., 2017) into a mixed lithology target (Fel'dman and Sazonova, 1993). The resulting impact formed a central uplift region of roughly 5.5 km in diameter and an *apparent* outer rim diameter of about ~13 km (Masaitis *et al*., 1984). *Zhamanshin* is moderately eroded with widespread loess deposits further obscuring impact-related morphologies (Florenskii and Dabizha, 1980; Garvin et al., 1992; Indu et al., 2022). The impact structure is minimally vegetated (see **Appendix 1** *field photos*).

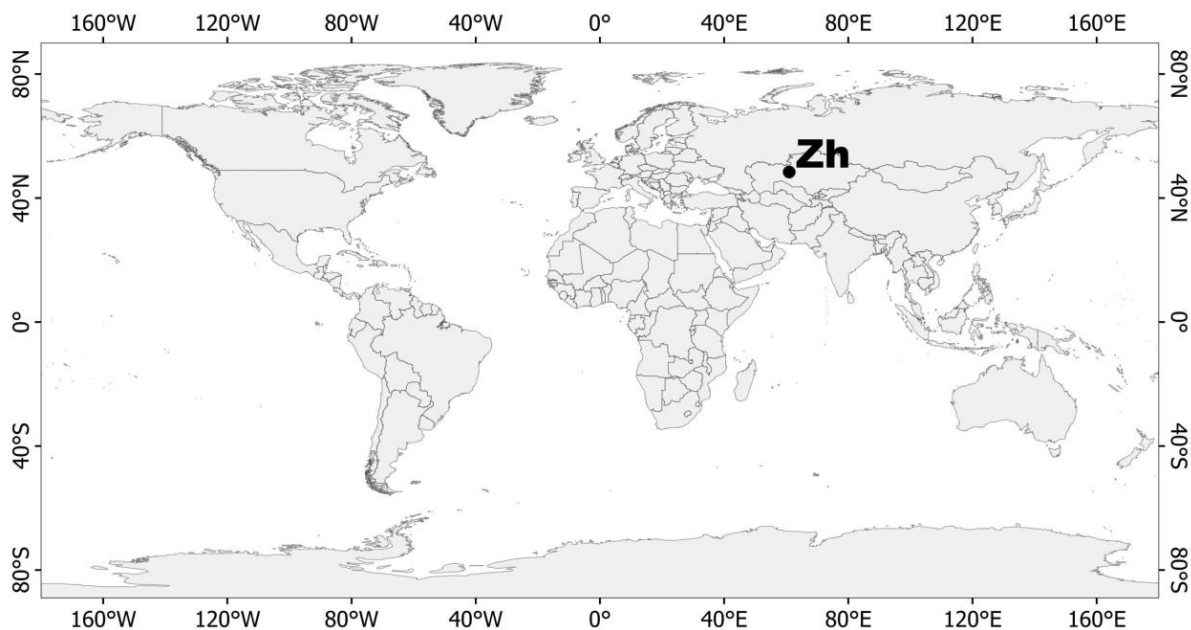

**Figure 1**. Location of the (**Zh**) *Zhamanshin* impact structure in Kazakhstan. Field images by the authors of the physiography of this region are illustrated within **Appendix 1**.

*2.2 Digital Elevation Models*

Five DEM datasets were used in this study (**Table 1**), including: 50 m gridded GEDI laser altimetry (Dubayah et al., 2022), 30 m NASADEM (NASA JPL, 2021), 12 m TanDEM-X (Krieger et al., 2007), 4 m Planet-derived DEM (Noh and Howat, 2026), and 2 m EarthDEM (Noh and Howat, 2017; Porter et al., 2024). The NASADEM, TanDEM-X, Planet-derived DEM, and EarthDEM data sets are digital surface models and represent the elevations of all surface features. The gridded GEDI DEM is a digital terrain model and represents the *bare Earth* elevations void of any surface features such as vegetation with sub-meter vertical accuracy (Dubayah et al., 2022). However, *Zhamanshin* is minimally vegetated making this distinction minor. The term DEM is used interchangeably with digital surface model and digital terrain model herein. These DEM datasets represent the highest horizontal and vertical resolutions available at the *Zhamanshin* impact structure. Further description of the *Planet*-derived DEM dataset is provided in **Appendix 2**.

**Table 1.** Digital Elevation Models Used in this study.

| Name | Elevation Retrieval Method | Data Source | Horizontal Resolution (m) |
|---|---|---|---|
| EarthDEM | Stereogrammetry | WorldView-1, WorldView-2, WorldView-3, and GeoEye-1 | 2 |
| Planet-derived DEM | Stereogrammetry | PlanetScope Dove Classic | 4 |
| TanDEM-X | Radar Interferometry | TerraSAR-X | 12 |
| NASADEM | Radar Interferometry | Shuttle Radar Topographic Mission | 30 |
| Gridded GEDI | Laser altimetry | Global Ecosystem Dynamics Investigation | 50 |

**Note**s. The Gridded GEDI DEM provides better than ~30 cm vertical accuracy based on cross-over analysis by the GEDI team.

*2.3 Radial Profile Analysis System Algorithm*

We applied the Radial Profile Analysis System (Anderson et al., 2025) to each DEM data set to determine the presence of a possible apparent outer rim outside of the most recently published ~13 km apparent outer rim (Osinski et al., 2022) (UWO Earth impact database accessible online at https://impact.uwo.ca/). The Radial Profile Analysis System used a DEM as an input to perform a computational assessment of putative topographic rings (**Figure 2**). First, radial elevation profiles stemming from the impact structure centroid were computed across a full range of azimuths at 1-degree steps (**Figure 2, B-D**). Each radial elevation profile was detrended to remove regional tilts and ramps. Topographic peaks were identified within each radial elevation profile and their distance from the impact crater center was cataloged (**Figure 2C, E-H**). The distance a topographic peak was from the impact crater center was then used to identify topographic peaks at similar distances around the impact crater to determine the presence of a topographic ring. Detection of topographic rings was completed across multiple search iterations (**Figure 2, I-L**). Search iterations varied by maximum search distance from the impact crater centroid, minimum distance between topographic peaks within the same radial elevation profile, and the peak matching window size for matching topographic peaks between radial profiles. The total number of search iterations depended on the DEM horizontal resolution. More search iterations were used for DEMs with higher horizontal resolutions compared to DEMs with lower horizontal resolutions. For example, the 50 m x,y gridded GEDI DEM used 240 search iterations while the 2 m EarthDEM (**Figure 4**) used 6,000 search iterations. This was due to the smaller step sizes with the 2 m EarthDEM when iterating through values for the minimum distance between topographic peaks within the same radial profile. The topographic ring with the smallest variance in detection frequency between all search iterations was selected as the most likely apparent outer rim diameter (**Figure 2M**). Ties when selecting the most likely apparent outer rim diameter were resolved by selecting the ring with the smallest diameter to emphasize conservatism given plausible geologic variability and associated uncertainties. Potential outer rim diameters were provided in ranges of 400 m due to the binning used to combine results across the four search distance iterations. See Anderson et al. (2025) for a complete description of the Radial Profile Analysis System and its validation on a range of terrestrial impact features.

An estimate of the inner ring diameter and centroid coordinates were required to initialize the Radial Profile Analysis System. Inner ring refers to the internal ring of hills rising above the crater floor of peak-ring impact structures (Turtle et al., 2005). A best-fit circle was computed for the inner ring of *Zhamanshin* using three manually selected points along the inner ring relief crest. This process was repeated four times while selecting different points along the rim crest each time to compute a best-fit circle. The average diameter and centroid coordinates were calculated from the four best-fit circles to be used as input for the Radial Profile Analysis System. An average inner ring diameter of 12.5 km was calculated, which falls within the 12-13 km range of current best apparent outer rim diameter for Zhamanshin. The values calculated from the four best-fit circles were used as input for each DEM type.

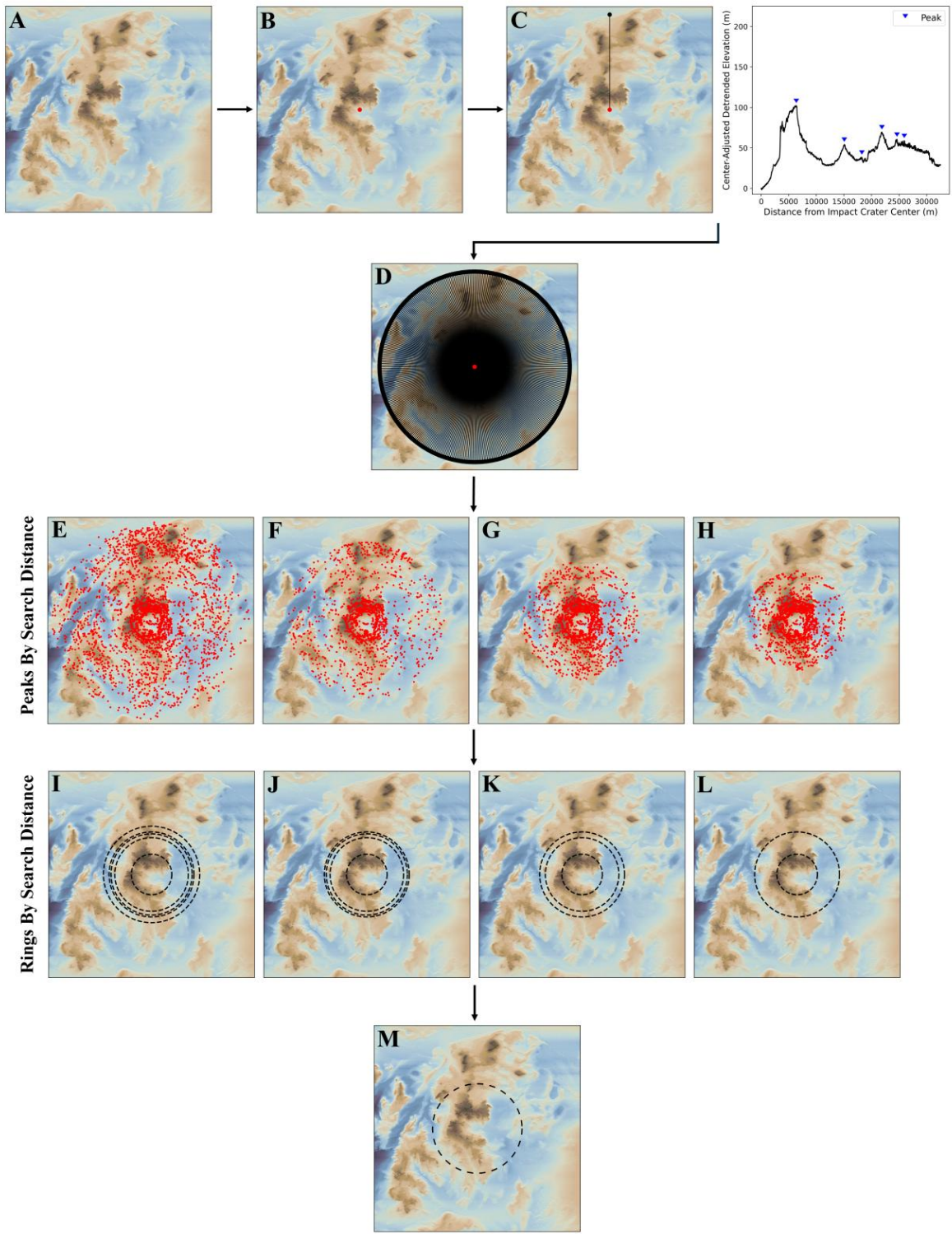

**Figure 2**. A graphical representation of the Radial Profile Analysis System over the *Zhamanshin* impact structure. Three inputs were required: (1) A DEM of the target impact crater (**A**); (2) An estimate of the crater centroid in UTM coordinates (red dot) (**B**); and (3) An initial estimate of the inner ring diameter. The centroid coordinate acts as an anchor point from which the Radial Profile Analysis System initiates

sampling (**C**). DEM elevations are sampled along a transect (black line) to construct an elevation profile. Transect length is computed using the inner cavity diameter estimate. Sampling and computation of the elevation profiles occur at each degree (of azimuth) in a full range of azimuths (0-359°) (**D**). Local maxima (red dots), or peaks, are identified within each elevation profile across four different search distances (in meters) from the crater centroid (**E-H**). Search distances are calculated using the inner ring diameter estimate within the following equations: 1) Inner Ring Diameter Estimate*1.1+2000 (**E**); 2) Inner Ring Diameter Estimate *2.5 (**F**); 3) Inner Ring Diameter Estimate *2 (**G**); and 4) Inner Ring Diameter Estimate *1.5 (**H**). Elevation profiles are subset to match the search distance and detrended to remove regional ramps and tilts before the RPS initiates its search for local maxima. Potential solutions for the inner ring diameter and outer rim diameter are computed using the identified local maxima (**I-L**). Potential rim solutions for each search distance are combined, and a final apparent outer rim solution is calculated from the aggregate pool of potential rim diameters (**M**). Plots were produced using TanDEM-X derivative products, which includes copyrighted material of Airbus Defense and Space GEO Inc., U.S. All rights reserved.

*2.4 Earth Impact Effects*

Environmental effects from the predicted apparent outer rim diameter were computed using the Earth Impact Effects Program on-line system (Collins *et al*., 2005; Available from: impact.ese.ic.ac.uk). A range of impactor diameters, densities, velocities, angles of impact, and target properties were iterated through to identify the parameters that result in the apparent outer rim diameter predicted by the Radial Profile Analysis System (i.e., 26-27 km). Parameter ranges were constrained based on information provided in the extant literature (Schmieder and Kring, 2020). The best-fit parameters were then used to estimate the possible environmental effects from the resulting *Zhamanshin* impact event at ~0.87 Ma.

**3. RESULTS**

3.1 Apparent Outer Rim Prediction

Predicted apparent outer rim diameters for each DEM type are provided in **Table 2** and shown graphically in **Figure 3**. The Radial Profile Analysis System predicted an apparent outer rim diameter of 26.4-26.8 km for each DEM when using an inner cavity diameter estimate of 12.5 km. No other possible apparent outer rim diameters were identified. Uncertainty in predicted apparent outer rim diameters varies by DEM data and was found to be between 0.6-4.1 km on average (Anderson et. al., 2025). Prediction uncertainty has not been quantified when using the 50 m gridded GEDI DEM. The apparent outer rim diameters predicted here result in a change upward of 13.4-13.8 km from the current best estimate (Burba, 1994; Florenskii and Dabizha, 1980; Garvin and Schnetzler, 1992; Schmieder and Kring, 2020), or an increase of 103-110%. **Figures 4 and 5** show the predicted 26.4-26.8 km outer rim diameter overlaid on the highest spatial resolution DEM of *Zhamanshin* (2 m EarthDEM) available (**Figure 4**). Additional parameters were explored (**Appendix 3** and **Appendix 4**).

**Table 2.** Radial Profile Analysis System algorithm predictions for the *Zhamanshin* impact structure by DEM type.

| DEM Type | Horizontal Resolution (m) | Predicted Apparent Outer Rims (km) |
|---|---|---|
| Gridded GEDI | 50 | 26.4-26.8 |
| NASADEM | 30 | 26.4-26.8 |
| TanDEM-X | 12 | 26.4-26.8 |
| Planet-derived DEM | 4 | 26.4-26.8 |
| EarthDEM | 2 | 26.4-26.8 |

**Notes.** See text for additional details and **Supplemental Section S3**.

To visualize the topology of the newly estimated outer tectonic rim (via Radial Profile analysis System), an oblique "ray-trace" view of the *Zhamanshin* structure is illustrated in **Figure 5**, with the inner ring (12.5 km) in a solid red circle (inner structural uplift region) and the outer rim in dashed red bars. This showcases the breaks in slope that can be observed in ground photographs (**Appendix 1**) where there is an absence of any field geophysical survey datasets precluding independent assessment as recently reported by Quesnel and others (Quesnel et al., 2024).

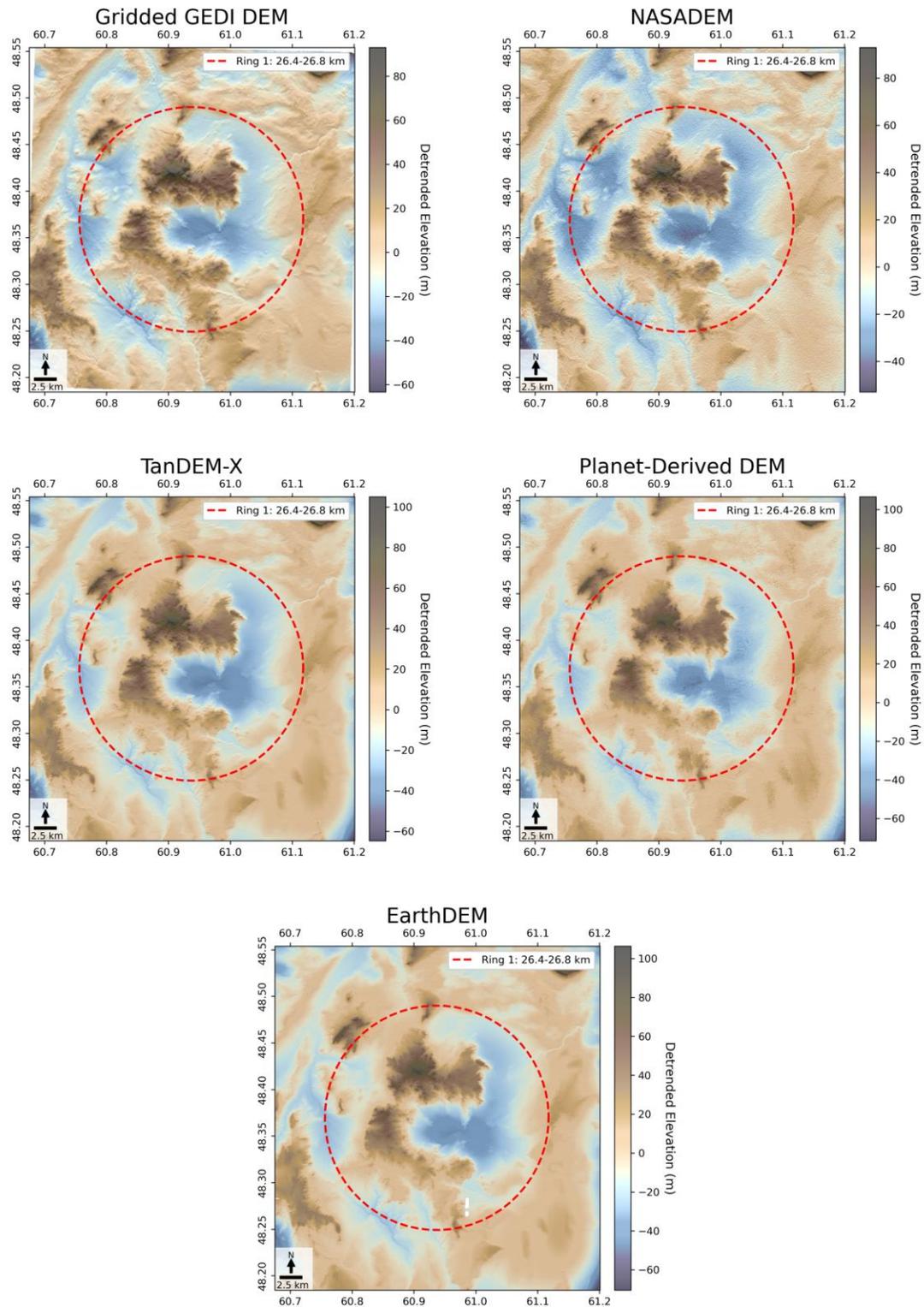

**Figure 3.** Apparent outer rim diameter predicted by the Radial Profile Analysis System for *Zhamanshin* when using the 50 m gridded GEDI bare-Earth DEM, 30 m NASADEM, 12 m TanDEM-X, 4 m Planet-derived DEM, and 2 m EarthDEM. The most likely apparent outer rim diameter as determined by the Radial Profile Analysis System algorithm is shown in **red**. The Radial Profile Analysis System algorithm

identified a single apparent outer rim diameter for each of the five DEMs. See **Appendix 2** for additional details.

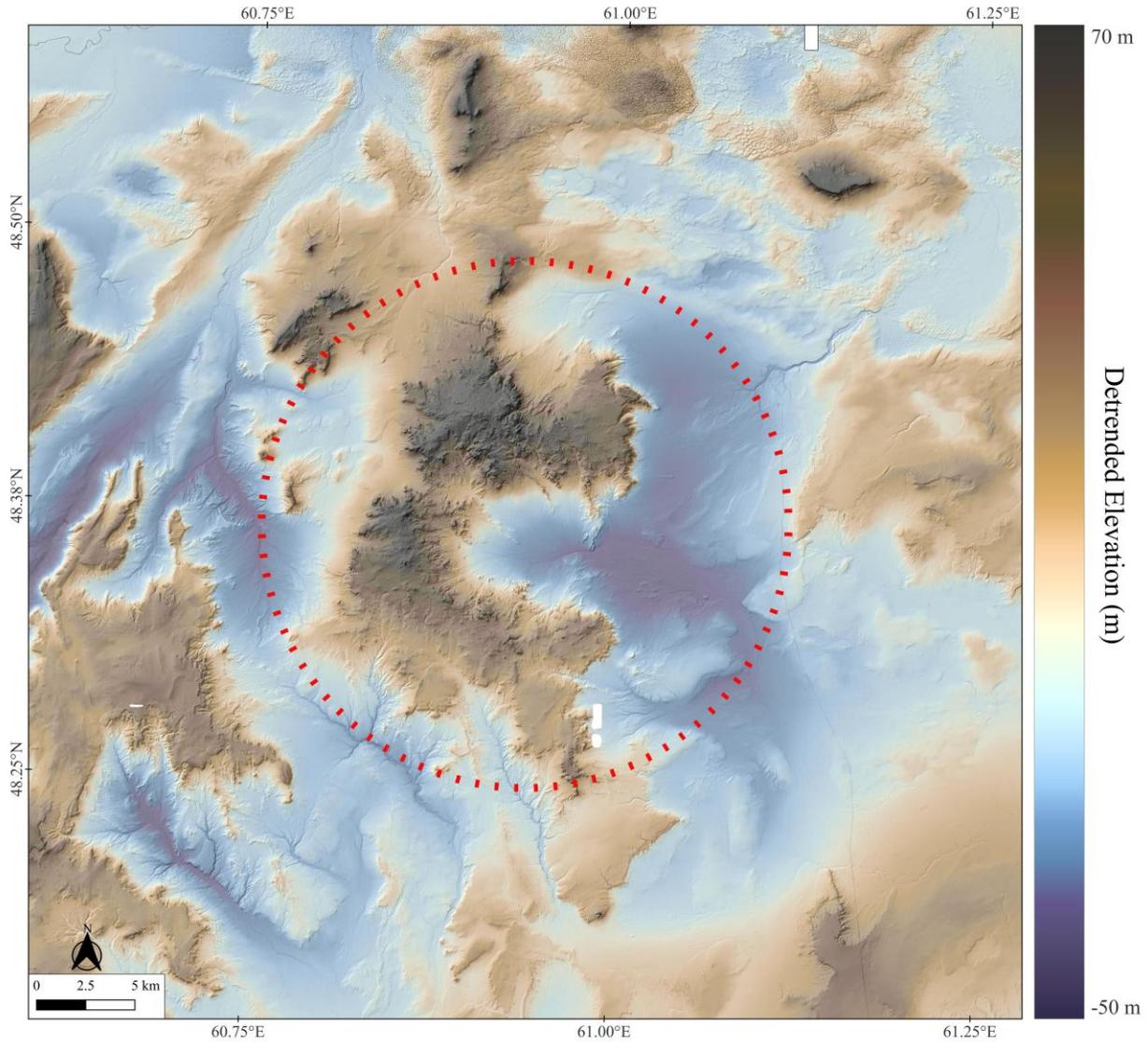

**Figure 4.** Highest spatial resolution surface elevation data (2 m x,y) EarthDEM of the *Zhamanshin* area with a dashed **red** predicted outer ring as the putative tectonic rim with width of the red dashes representing the range of solutions (~400 m long). Color scale covers a ~120 m range of local topography, with an apparent inner ring at 12.5 km (not shown). Note breaks in slope in the high-resolution DEM with termination of drainages and associated features around the location of the putative outer rim at ~26.4 km as discovered via the Radial Profile Analysis System. See **Appendix 1** for ground views of regions where the red dashes denote expression of an outer rim.

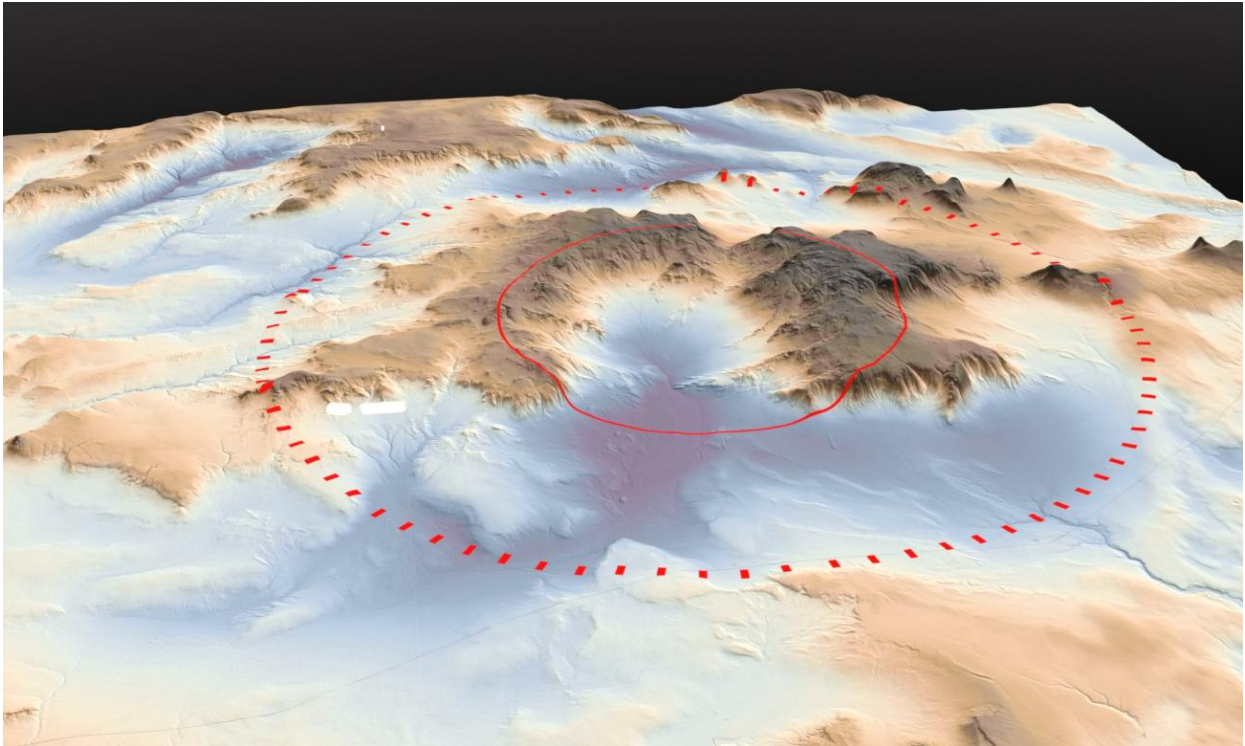

**Figure 5**. Oblique perspective view from southeast of the 2 m EarthDEM dataset (**Figure 4**) showing the topology of the impact structure with the 26.4-26.8 km predicted outer tectonic rim in **red** dashes. The width of the red dashes represents the 400 m prediction interval. The solid **red** line indicates the 12.5 km inner ring diameter. Note the correlation of drainage directional changes with the dashed red outer ring computed via the Radial Profile analysis System in this view.

3.2 Environmental Effects

Two "endmember" cases were identified for the *Zhamanshin* impact structure that resulted in a final crater diameter of ~26 km (**Table 3**) using the EIE Program (Collins et al., 2005) with additional considerations from extant literature (Melosh, 1989; Osinski et al., 2022). **Case 1**, an impact into a crystalline target with a final crater diameter of ~26 km, would have released approximately 440,000 Mt of kinetic energy (Collins et al., 2005). This would have caused regional environmental effects from 100-1500 km from the impact point. Roughly 11.2 km$^3$ of target material would be melted or vaporized with half of the material being displaced beyond the primary complex crater. The air blast would have traveled ~1500 km from the impact center in ~ 80 minutes with a maximum wind velocity of ~17 mph. A fine dusting of impact sediments, ~ 180 mm in thickness, would have reached 1500 km from the impact center about 10.6 minutes after impact. **Case 2**, an impact into a sedimentary target with a final crater diameter of ~26.5 km, would have released approximately 240,000-300,000 Mt of kinetic energy. Like an impact into crystalline target material (Case 1), this would cause regional environmental effects. Roughly 10.3 km$^3$ of target material would be melted or vaporized with half of the material being displaced beyond the primary crater. An air blast would reach ~1500 km from the impact center in roughly 80 minutes with a maximum wind velocity of ~17 mph. A fine dusting of impact sediments, ~ 500 mm in thickness, would reach 1500 km from the impact center about 10.6 minutes after impact. In both cases, the effects are consequential for regional environments due to widespread fires, loss of sunlight, and acid rain (Toon et al., 1997). Published estimates for the time for recovery from any major impact event with more than 240,000 Megatons kinetic energy is estimated to be in the range of thousands to up to a million years (Osinski et al., 2022; Toon et al., 1997).

**Table 3**. Earth Impact Effect (EIE) program parameters used to calculate possible environmental effects for two endmember cases.

| Endmember Case | 1 | 2 |
|---|---|---|
| **Projectile Diameter (m)** | 1,500 | 1,500 |
| **Projectile Density (kg/m$^3$)** | 3,000 | 3,000 |
| **Impact Velocity (km/s)** | 25 | 25 |
| **Impact Angle (degrees)** | 45 | 45 |
| **Target Type** | Crystalline | Sedimentary |
| **Target Density (kg/m$^3$)** | 2,750 | 1,500 |

**Notes.** Both cases result in a final crater diameter like the predicted 26.4-26.8 km apparent outer rim diameter identified at the *Zhamanshin* impact structure.

These are endmembers and as kinetic energy is strongly related to impact velocity to the second power and to the density of the impactor. Thus, one can evaluate impactor velocities from 17 km/s to over 30 km/s for different densities of impactor (at a fixed impactor diameter of ~1.5 km) to achieve kinetic energy values over 440,000 Mt while maintaining the complex multi-ringed impact crater final diameter between 26-27 km. Energy release of ~ 240,000 Mt is 1200 times greater than the *Krakatau* volcanic explosion of 1883 (200 Mt) and 16,000 times more energetic than the 1908 *Tunguska* event, but only 0.33% as much as the K-Pg *Chicxulub* impact (i.e., 10-12 km projectile at 66 Ma [Morgan et al., 2022]). An impact event resulting in ~240,000-500,000 Mt of kinetic energy could result in serious environmental effects (EIE program; Toon et al., 1997), including some of the following specific shorter-term examples:

- Short-term environmental impact of ~ 2-3 degrees C "impact winter" for 3-4 years; localized temperature drops up to ~10 degrees C for about a month as the impact-induced dust cloud (fines < 1 mm) disperses globally enabling a return to pre-impact levels (Osinski et al., 2022; Toon et al., 1997).
- Global acid rain precipitation for several years.
- Depletion of ozone in the stratosphere for several years triggered by global injection of dust and aerosols (and regional to global distribution of tektites).
- Increased $CO_2$ production for ~ 10 years contributing to the greenhouse gas effect.
- Regional fireball up to ~100 km radially from the point of impact.

## 4. DISCUSSION

Reducing the uncertainties in the critical spatial dimensions of the most recent complex terrestrial impact cratering events (≤ 1 Ma) requires an Earth-centric approach using newly available data and innovative methods. This study used high spatial resolution DEMs and the Radial Profile Analysis System (Anderson et al., 2025) to reassess the spatial dimensions of *Zhamanshin* impact structure. Our evaluation of multiple high resolution DEM data sets suggests that the current best estimates for the apparent outer rim diameter listed in the extant literature for *Zhamanshin* could be uncertain by 13.4-13.8 km (**Table 1; Figure 3**). Topographic features fitting the Radial Profile Analysis System's criteria to be considered an impact-related topographic ring were consistently identified 13.2-13.4 km from *Zhamanshin*'s centroid. These features were identified when using DEMs derived using different (and independent) methods for estimating surface elevations (stereo-photogrammetry, InSAR, and geodetic laser altimetry), horizontal resolutions (2-50 m), and vertical accuracies (0.30-15 m) (Dubayah et al., 2022; Uuemaa et al., 2020). Despite these findings, additional surveys are clearly required to validate the apparent outer rim diameter predicted by the Radial Profile Analysis System. Anderson and others (2025) determined the Radial Profile Analysis System's average error in apparent outer rim diameter for terrestrial impact structures to be 0.6 - 4.1 km depending on the DEM data set utilized. It is suggested that future studies of the *Zhamanshin* impact structure take the Radial Profile Analysis System measured errors into consideration when defining the spatial extent of field and airborne geophysical surveys.

4.1 Implications of larger outer rim diameter

      An apparent outer rim diameter of 26.4-26.8 km (**Figure 5**) is a significant discrepancy (Grieve, 2017; Turtle et al., 2005) when considering the consequences associated with the plausible kinetic energy of impact magnitudes (Collins et al., 2005; Davison and Collins, 2022) for this class of ultra-recent complex cratering event. A factor of 7.5 to 10 increase in the likely kinetic energy is plausible when using the apparent outer rim diameter predicted by the Radial Profile Analysis System discussed herein. This corresponds to more than 240,000 Mt released into the Earth's atmosphere-land system compared to 39,000 Mt computed using an apparent outer rim diameter of ~13 km. The effects from more than ~240,000 Mt into the Earth's environmental system would be present for 10-1000's of years (i.e., including cool down and global dust distribution ala Grieve and Shoemaker (1994) and Pierazzo and Artemieva (2012)). The most recent analogue to the *Zhamanshin* impact event is the *Ries* impact crater of Germany, which formed at ~14.8 Ma and has an apparent outer rim diameter of ~26 km (McCall et al., 2024; Schmieder and Kring, 2020). The estimated environmental effects of the *Ries* impact event included evidence of obliteration of life within a 100 km or larger radius with global consequences (EIE program iterations & Pierazzo and Artemieva (2012)). Thus, the consequences of the *Zhamanshin* impact event at a time of an already-established climate variability during the early Middle Pleistocene Transition (Clark et al., 2006; Indu et al., 2022) should be considered due to the possibility of preserved signatures of such large explosive events and their consequences (i.e., globally distributed stratospheric dust, induced regional mega-tsunamis with associated deposits, *etc*.; *see for example* Osinski et al., (2022) and Pierazzo and Artemieva (2012)). It is therefore provocative to speculate whether a cratering event of this scale could have contributed to the global environmental adjustments at the *Middle Pleistocene Transition*, just as those associated with paroxysmal volcanic eruptions (e.g., *Yellowstone* Caldera at kinetic energy of ~ 875,000 Mt [Christiansen, 2001; Perkins and Nash, 2002]) are correlated in time with similar scale effects at other times (Toon et al., 1997). These critical correlations are yet to be accomplished but represent an important next step in this trajectory of research, with astrobiological consequences beyond Earth.

      This research demonstrates that meaningful reductions in uncertainties of the apparent outer ring (as a tectonic rim) diameters of complex impact features are plausible now due to the availability of high spatial and vertical precision DEMs, which serves as a reliable boundary condition for estimating key dimensions (diameter, depth, model-based kinetic energy of impact), as previously recommended by Turtle et al. (2005). Prior assessments relied upon partially subjective evaluations from 2D datasets (imaging) with sometimes error-prone outcomes (Ait Oufella et al., 2025; Turtle et al., 2005). The approach used in this study provides an Earth-centric method for the quantitative analysis of high-resolution digital topography from which a range of potential apparent outer rim diameters can be computed. Results from *Zhamanshin* advocate for the further examination of other very recent NEO events, such as *Bosumtwi* and *Pantasma*. Resolving dimensional uncertainties in terrestrial complex impact events can additionally improve continental erosion rates (Perron, 2017) as recently developed by Hergarten and Kenkmann (2019) and elucidate variabilities across multi-million-year intervals. Further evaluation of the critical geometric properties of *Zhamanshin*, as discussed above, is essential, given the implications of these large revisions based on newly available DEM datasets and should include field-based assessment ideally with both geological sampling and *wider-area* geophysics (gravimetry, as noted in Quesnel et al.,2024).

4.2 Connection to Mars

      With increasing interest in both scientific and human-based exploration of Mars, it is instructive to consider whether any generally "kinetic energy equivalent" complex impact structures less than ~ 5 Ma in apparent age can be identified on the planet Mars. Using classical scaling relationships (Collins et al., 2005; Croft, 1985; Kundu et al., 2019; Melosh, 1989), a ~ 26 km terrestrial impact structure such as

*Zhamanshin* would be over ~50 km in final diameter on Mars based on lower gravitational acceleration and other factors. Recent studies on the Martian impact record have identified several ultra-recent complex craters (Goddard et al., 2014; Werner et al., 2014; Williams and Malin, 2008) that could be compared to the 0.87 Ma *Zhamanshin* event on Earth, including the ~ 58 km *Mojave* crater (7.5N, 33W) and the ~ 39 km *Kotka* crater (18.5N, 171.2W).

We applied the Radial Profile Analysis System to the 200 m MOLA-HRSC Blended DEM (Fergason et al., 2018) of *Mojave Crater* as a preliminary experiment. Initial results suggest an outermost tectonic rim at ~60-64 km, slightly larger than 2D image-based estimates (**Figure 6b**). Simple application of the principles employed in the Imperial College *Earth Impact Effects* program (Collins et al., 2005) with extension to Mars (via different gravity, atmosphere, etc.) suggests that an asteroidal impact on Mars capable of creating a ~ 60 km complex crater would result in at least 500,000 Mt of impact kinetic energy, large enough to generate global consequences for the Martian troposphere and distribution of dust over thousands of years. Goddard and others describe recent evidence of fluvial activity at *Mojave* (Goddard et al., 2014) perhaps connected to the melting of buried ice and release of ground water and potentially impact-induced pluvial activity. If the *Mojave* impact event on Mars excavated equatorial ground ice and produced a short-term release of surface water and rainfall, some of the pristine landforms described by Williams and Malin (2008) and others could be explained. A short term period of enhanced habitability with surface and near-surface groundwaters and runoff, plus a thermal anomaly lasting tens of thousands of years (Pierazzo and Artemieva, 2012) would have obvious astrobiological interest, as well as serving as a useful site for future in situ resources utilization (ISRU) applications potentially in support of NASA's Moon to Mars initiative as outlined in the 2025 President's Budget Request (PBR, 2025). Perhaps *Zhamanshin* can shed insights into the longer-term evolution of the geologically recent *Mojave* crater on Mars, as consideration of future robotic and human landing sites are advanced over the next decade.

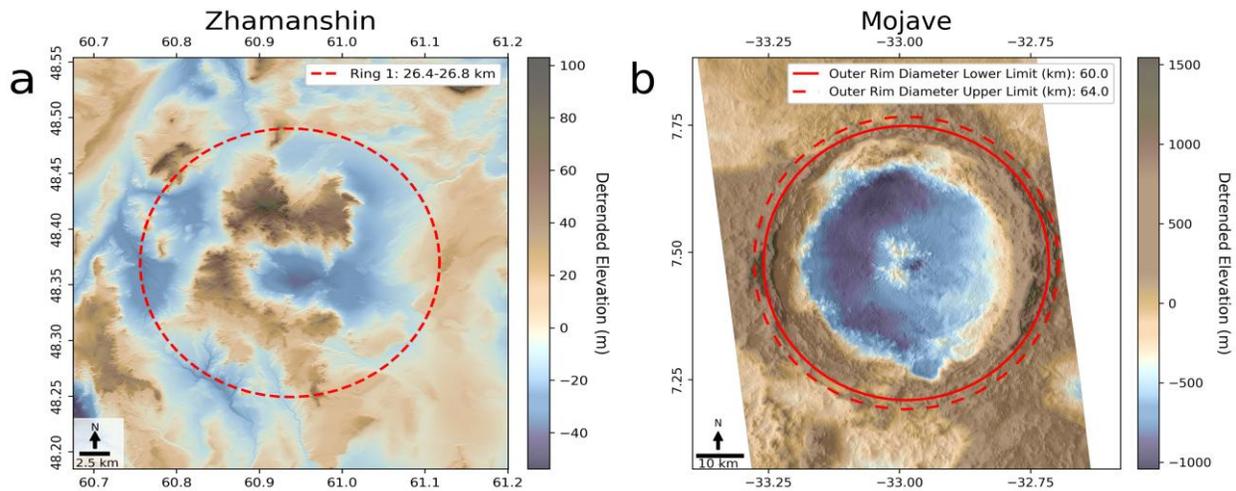

**Figure 6.** Radial Profile analysis System results for *Zhamanshin* (**a**) using TDX DEM (12 m x,y) compared to the Martian complex impact structure *Mojave* (**b**) using the 200 m MOLA-HRSC Blended DEM (Fergason et al., 2018). The *Mojave* results are overlaid on a ~10 m DEM created from 9 overlapping MRO CTX images for improved viewing of surface topography at a similar scale to that shown for *Zhamanshin* in (**a**). The 10 m CTX-derived DEM was *not* processed using the Radial Profile Analysis System due to incomplete coverage of surrounding ejecta area to the west and east of the crater's cavity.

## 5. CONCLUSIONS

Our quantitative analysis of five DEM data sets indicates that the *Zhamanshin* impact structure could be twice as large as previously thought with an estimated ~ 240,000-500,000 Mt total range of kinetic energy computed from the revised apparent outer rim diameter. This far exceeds the threshold required for limited atmospheric blowoff (Grieve, 2017; Toon et al., 1997). The environmental consequences of a ~ 26.5 km diameter impact cratering event in the ~ 870,000 (± 20,000) yr (BCE) time interval of Earth history (*early* Middle Pleistocene Transition) have not previously been considered. An impact event of this magnitude could have triggered global environmental effects (Osinski et al., 2022; Pierazzo and Artemieva, 2012) that resulted in unknown biological responses at a variety of spatiotemporal scales (e.g., Ashton and Stringer, 2023; Hu et al., 2023; Margari et al., 2023; Voosen, 2023). The duration of the post impact events and whether they were of consequence to regional (or beyond) biological systems is not known nor established at this time. However, the duration of post impact consequences can be constrained to be decades at minimum based on the stratospheric injection of gas, dust, and impact glass (tektites) and their typical residence times (Collins et al., 2005; Pierazzo and Artemieva, 2012).

Further analysis of climate records in the interval from ~0.8 to 0.9 Ma for evidence of specific perturbations in the preserved photosynthesis record (including palynology) is one recommended next step in response to the revised diameter of the *Zhamanshin* impact structure. The terrestrial impact record for the past one million years is of known consequence to the evolution of biological systems on Earth and additional efforts to unravel its incomplete history are compelling, considering key goals associated with planetary protection of Earth (e.g., Garvin et al., 2023; Pierazzo and Artemieva, 2012; Schmieder and Kring, 2020). Reevaluation of additional well-preserved and recent hypervelocity impact structures with high-resolution digital topography will continue to support a reduction in uncertainties of key parameters (Turtle et al., 2005) associated with these environmentally relevant exogenic events within Earth history, and potentially also for Mars (Werner et al., 2014).

The revised (upward by a factor of two) values for key spatial dimensions of *Zhamanshin* reported herein as part of ongoing efforts (Garvin et al., 2023) to reduce uncertainties in dimensional parameters of large-scale terrestrial impact events, has a direct bearing on preserved environmental (and related climate) records on Earth in a time interval during which noteworthy evolutionary changes have been suggested (Clark et al., 2006; Perron, 2017). These initial results, while suggestive, require follow-on research in the form of regional geophysical surveys like those conducted at complex impact structures such as *Haughton* (Grieve, 1988) and others including the *Ries* impact event at ~ 14.8 Ma (McCall et al., 2024). New boundary conditions for additional well-preserved, very recent hypervelocity impact structures in the form of high-resolution topography and gravimetry will continue to support a reduction in uncertainties of key parameters associated with these environmentally relevant exogenic events within Earth history. Extending this work to the *most recent* multi-ringed complex impact events on Mars (in the > 39 km diameter size range as with *Mojave*) is another fertile area of research to be considered in an astrobiological context (e.g., preserved, recent biosignatures) and in support of NASA's Moon to Mars exploration priorities (PBR, 2025).

## ACKNOWLEDGEMENTS

*This work utilized data made available through the NASA Commercial Smallsat Data Acquisition (CSDA) Program. EarthDEM DEMs provided by the Polar Geospatial Center under NSF-OPP awards 1043681, 1559691, 1542736, 1810976, and 2129685. We gratefully acknowledge the support of the Planetary Defense Coordination Office (Dr. Kelly Fast), and from the ISS/GEDI Program Scientist, plus encouragement by Dr. Chris Scolese and Dr. Tom Wagner. Motivation for this work began with discussions involving Dr. Charles Schnetzler that commenced in 1985 and continued until his death in 2006. Long-standing motivational mentorship from Richard Grieve is also acknowledged. Special*

*thanks to Emil Izokh and others (V. Masaitis) for arranging the 1989 International expedition to Zhamanshin which produced key samples for radiometric age determination. Additional thanks to Dr. Daniel Slayback (SSAI at GSFC) for assistance with GIS analyses along the way as part of this DEM research. Special processing of ISS/GEDI data by J. Bryan Blair, M. Hofton, and M. Linkswiler over Zhamanshin at high vertical precision contributed to this research and is gratefully acknowledged.*

**Appendix 1. Field Photos of *Zhamanshin***

      Photos collected in September 1989 by James B. Garvin as part of first USSR-Int'l Expedition to *Zhamanshin* arranged by the USSR *Akad. Nauka*. Photos were taken with 35 mm Leica SLR with Kodak color film (Kodachrome) in Sept.-Oct, 1989 while on an expedition to Zhamanshin.

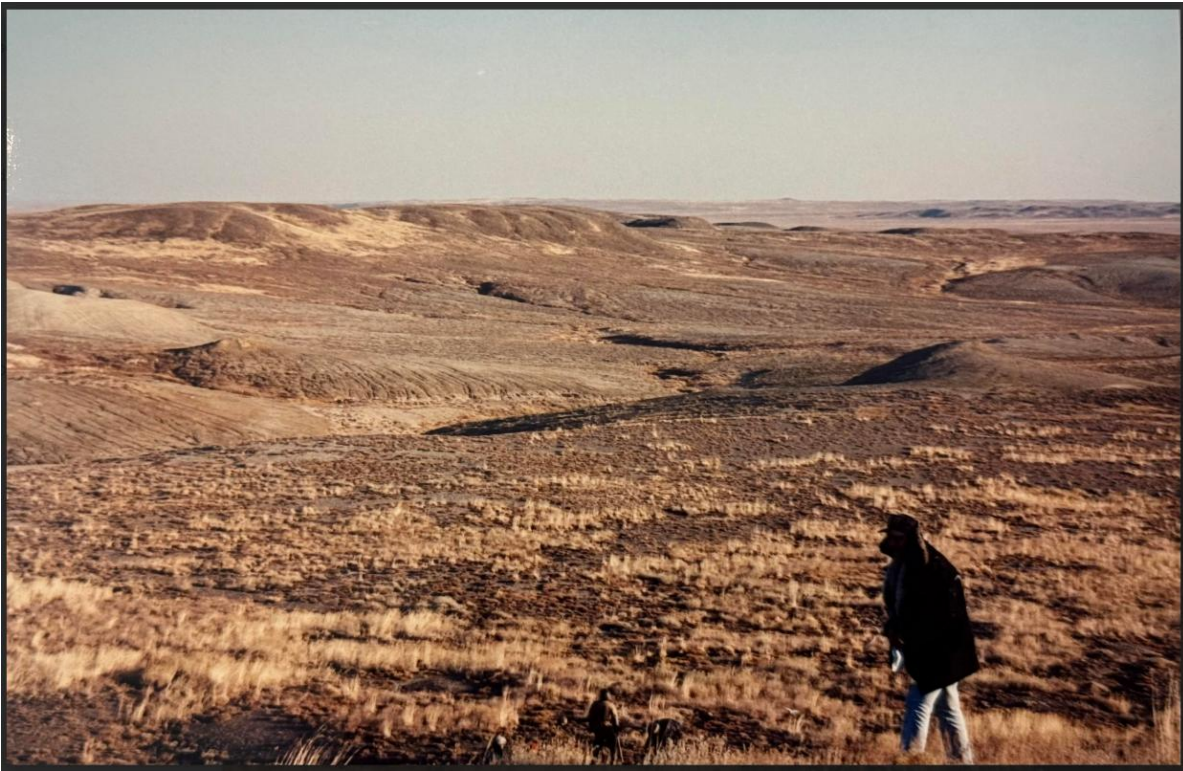

**Figure A1.** Interior of *Zhamanshin* with R.A.F. Grieve in foreground for scale with 10-20 m of local surface relief.

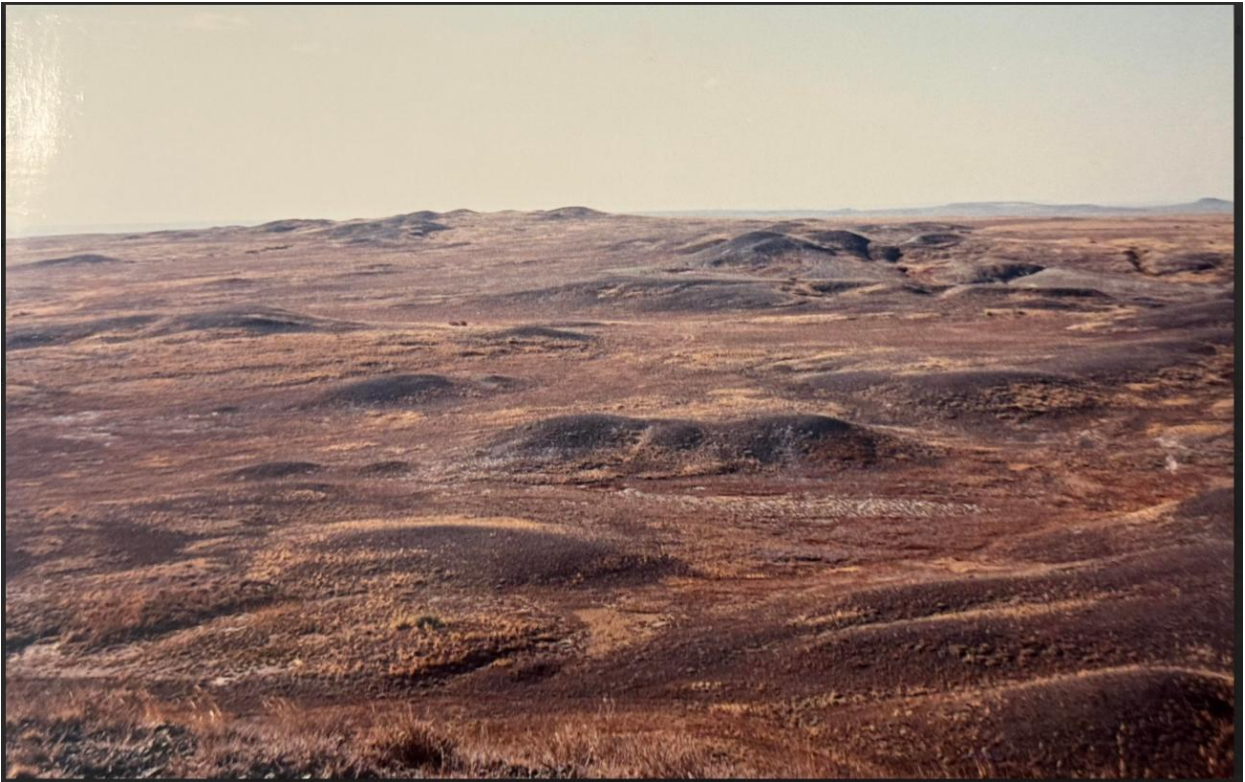

**Figure A2.** *Zhamanshin* outer region *hilly terrain* with several 10's of meters of relief looking toward NW from inner ring hills toward the Radial Profile Analysis System-identified outer tectonic rim.

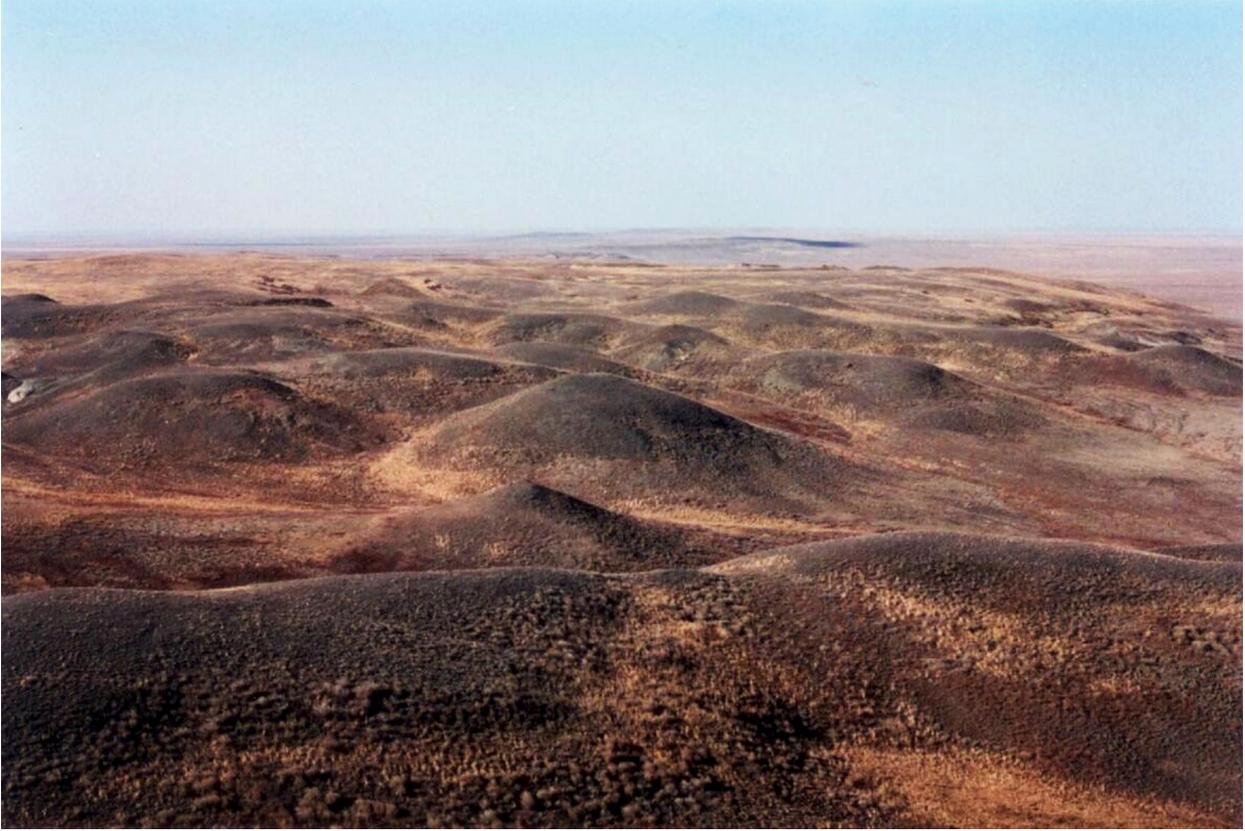
**Figure A3.** View at edge of rings of hills defining the 12-13 km inner ring looking to the West toward the Radial Profile Analysis System identified outer tectonic ring of hills at ~26.4 km (diameter).

**Appendix 2. Planet-Derived Digital Elevation Model**

We computed a DEM from multiple *PlanetScope* Dove Classic images using the methods described in Noh and Howat (2026). Generation of a DEM from *PlanetScope* Dove Classic images began with the selection of suitable stereo pairs from a set of input images. Our image set consisted of 1,182 images acquired between May 1, 2020 and June 30, 2020 (**Table S1**). All stereo pairs were then vertically co-registered to minimize inter-pair height inconsistencies and processed using the SETSM (Noh and Howat, 2025) multiple-pair image matching procedure with relative RPC-bias compensation. The output from this processing was a tiled DEM data set, which was then mosaicked to form a single DEM. It is important to note that the DEM computed from stereo pairs of *PlanetScope* Dove Classic images does not represent bare-earth elevations (as with GEDI), i.e., it is a digital surface model (DSM). The term DEM was used synonymously with DSM within this paper.

**Table A1.** PlanetScope Dove Classic Images Used to Create Planet-derived DEM of Zhamanshin

| Impact Crater | Number of Images | Date Range |
|---|---|---|
| *Zhamanshin* | 1,182 | May 1, 2020-June 30, 2020 |

**Notes.** Count and date range of *PlanetScope* Dove Classic images used to produce a DEM of the *Zhamanshin* impact structure (see **Figure 3**).

**Appendix 3**. **Predicted Apparent Outer Rim Diameter by Inner Ring Diameter Estimate**

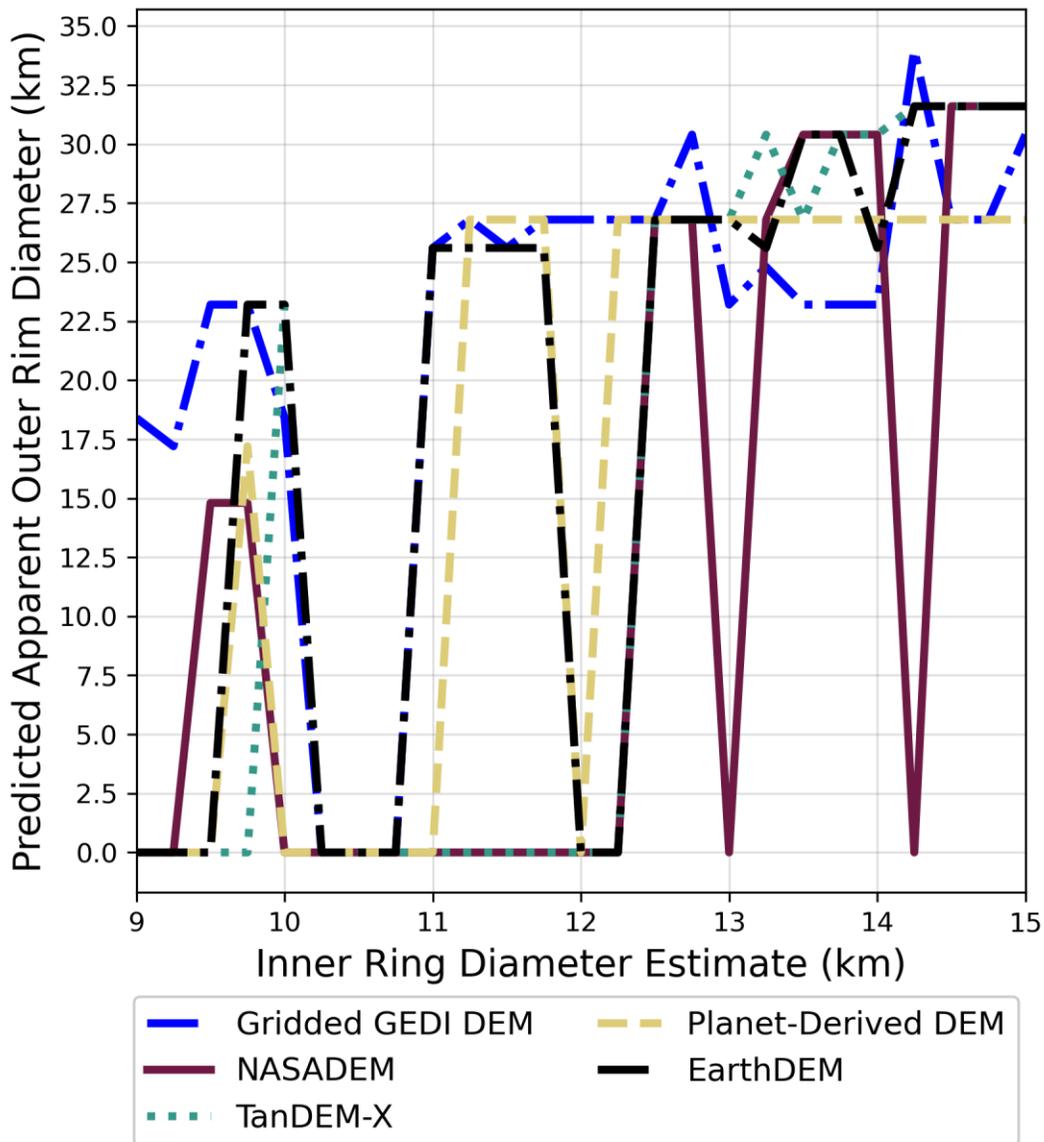

**Figure A4.** The predicted apparent outer rim diameter when iterating the inner ring diameter estimate for the *Zhamanshin* impact structure between 9-15 km with a radial step size of 250 m. The 12-13 km inner ring diameter matches the current best estimates for the apparent diameter. An apparent outer rim diameter of 26.4-26.8 km was reported for each DEM type with an inner ring diameter estimate of 12.5 km. The Radial Profile Analysis System reports diameters as a range, and the diameter reported in this figure represents the upper limit of that range. The lower limit of the range can be calculated by subtracting 400 m from the upper limit. A predicted apparent outer rim diameter of zero indicates that the Radial Profile Analysis System failed to identify an apparent outer rim.

**Appendix 4. Predicted Apparent Outer Rim Diameter When Varying the Minimum Radial Profile Count Required for a Topographic Ring**

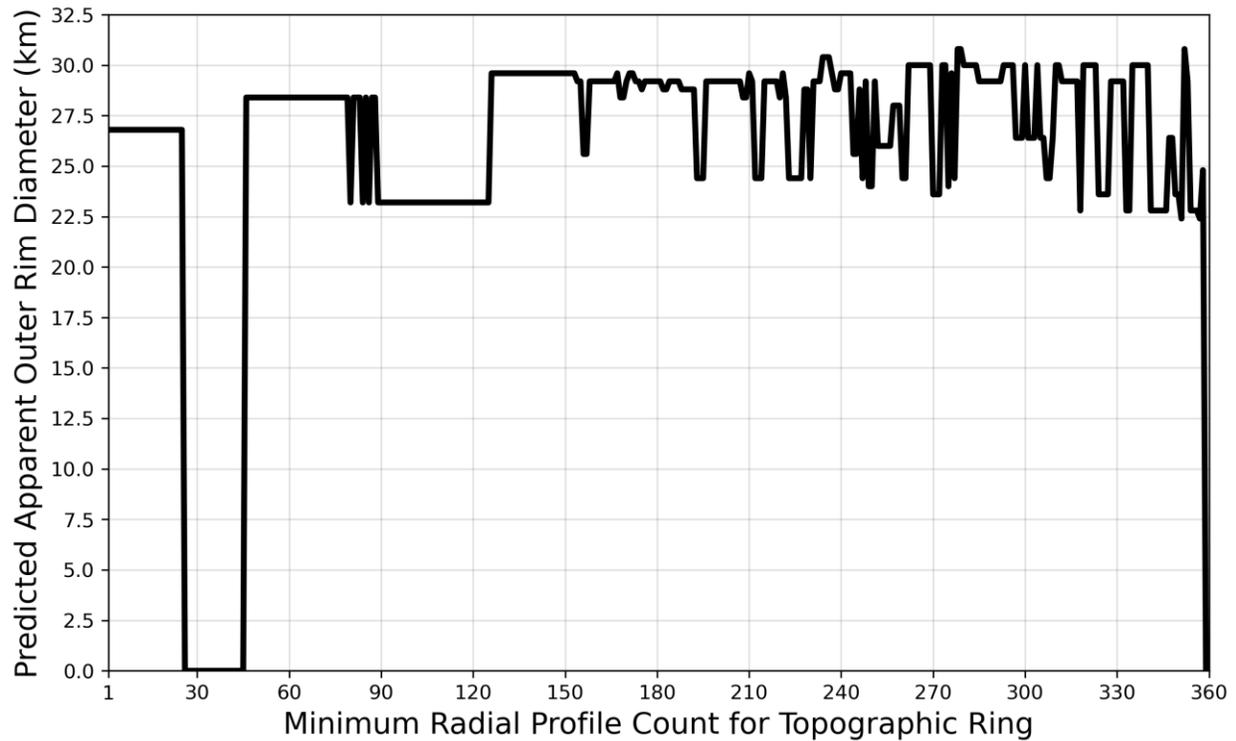

**Figure A5.** The most likely apparent outer rim diameter when iterating the number of radial profiles a topographic peak needed to be present within to be considered a topographic ring (Section 2.3). The minimum radial profile count was iterated between 1-360 with a step size of one. The default setting for the Radial Profile Analysis System was a minimum radial profile count of five. All iterations used an inner ring diameter estimate of 12.5 km. The Radial Profile Analysis System reports diameters as a range, and the diameter reported in this figure represents the upper limit of that range. The lower limit of the range can be calculated by subtracting 400 m from the upper limit. A predicted apparent outer rim diameter of zero indicates that the Radial Profile Analysis System failed to identify an apparent outer rim.